\documentclass[aps,prb,twocolumn,showpacs]{revtex4}
\usepackage{graphicx} 
\usepackage{epstopdf}

\begin{document} 

\title{Diffusion of Ga adatoms   at the surface of GaAs(001) c(4x4) $\alpha$ and $\beta$ reconstructions.} 

\author{Marcin Mi{\'n}kowski $^\dagger$ and Magdalena A Za{\l}uska--Kotur $^{\dagger,\ddagger}$}
\affiliation{$^\dagger$Institute of Physics, Polish
Academy of Sciences, Al.~Lotnik{\'o}w 32/46, 02--668 Warsaw, Poland \\
$^\ddagger$ Faculty of Mathematics and Natural Sciences,
Card. Stefan Wyszynski University, ul Dewajtis 5, 01-815 Warsaw, Poland }
\email{minkowski@ifpan.edu.pl, zalum@ifpan.edu.pl} 

\date{\today}

\begin{abstract}
Diffusion of Ga adatom  at the As rich, low temperature  c(4$\times$4) reconstructions of GaAs(001) surface is analyzed. 
We use  known   energy landscape for the motion of Ga adatom  at two different $\alpha$ and $\beta$ surface phases to calculate  diffusion tensor by means  of the variational approach.  Diffusion coefficient describes the character  of low density adatom system motion at the surface.  The resulting expressions   allow to identify main paths of an adatom diffusion and  to calculate an activation energy of this  process. It is shown that diffusion at  the  $\alpha$ surface is slower and more anisotropic than this for the $\beta$ surface. 
 \end{abstract}
\maketitle
\section{Introduction}
Surface diffusion is a control factor of layer by layer crystal growth, the one  of basic processes in the construction of nanotechnological   devices. The analysis  of  surface diffusion is  based on the determination of adsorption sites,  adparticle    binding energy  at each of  these sites and barriers for thermally activated jumps between them. When all these parameters are calculated, the next step is to describe particle diffusion process in this energy  landscape.
Development  of ab-initio calculations in the last years resulted in  very accurate description of the surface   energy  landscape at various  systems \cite{roehl,roehl2,shapiro,wheeler,ferron,gonzales,booyens,yi,podsiadlo,liu,antczak, antczak2,jalochowski,fichthorn}. For many crystals, important because of  their      applications in nanotechnology or biotechnology, full  energy map at the specific  surface orientations and reconstructions were done. The system which is  often a subject of study   is the  surface GaAs(001) in its various phases such as high temperature $(2\times 4)$ \cite{kley,13} or  low temperature $c(4 \times 4)$ \cite{13,etgens,resh,nagashima,ito,kaku,ohtake,nagashima2,penew,takahasi,iicek,ohtake2} reconstructions. This semiconductor  is very important in production of solar cells or microwave circuits used  in cellular phones. 
In this study we use presented in Refs. \onlinecite{roehl} and \onlinecite{roehl2} results of ab-initio calculations of energy landscape of As rich $c(4 \times 4)\beta$ and $c(4 \times 4) \alpha$ phases  for Ga adatom. On the base of these data we derive diffusion coefficient for Ga atoms and compare diffusive motion at both surface structures. 

The first observations at low temperatures and at the excess of As atoms established the existence of $c(4 \times 4)\beta$ structure  \cite{etgens,resh,nagashima,ito,kaku} and  then  asymmetric phase  $c(4 \times 4)\alpha$ \cite{ito,kaku,ohtake,nagashima2,penew,takahasi,iicek,ohtake2} was found. It is generally
accepted now that As-rich $ c(4 \times 4)$ surface is divided into two phases:
the $c(4 \times 4) \alpha$ phase (terminated by Ga - As dimers) which depending on the As pressure  exists up to 400- 550 K and the $ c(4 \times 4) \beta $
phase (terminated by As - As dimers) between 400K and 770K \cite{ito,13}.  Incorporation and then diffusion  of Ga atoms control growth process in  As rich conditions. That is why  detailed knowledge of the Ga adatoms diffusion is so important. Derivation of binding energies and  diffusion barriers is the  first step in such study. Further analysis can be based on some qualitative analysis of the energy structure \cite{roehl, roehl2}, Monte Carlo simulation data \cite{2,mattahai,ito,ind} or derivation of analytic formulas \cite{ind,
kley,haus}. 

 The analytic formulas have this advantage over other methods that they give results as a function of temperature and of all other model parameters.   We propose new, variational approach, that was first shown to work  for many-particle diffusion process \cite{GZ-K,Z-KG1,BZ-KG1,Z-KBG,Z-KG2}.  Below we show how the   variational approach can be used in calculation of tracer diffusion coefficient, which describes also diffusion in the   system of low density.  For systems of  low density  correlations can be  neglected. However the analysis of adatom motion is still not easy because   the reconstructed surface of  GaAs$(001) c(4\times 4)$ in both $\alpha$ and $\beta$ versions  contain many adsorption sites for   Ga adatoms and complicated  lattice of possible jumps  between them. Calculation of an effective diffusion coefficients in different directions and the analysis of possible modes of particle motion  is very difficult when  adatom diffuses over the surface of complex energy landscape.  It appears that on using proper variational analysis we end up with one formula for diffusion tensor, which then can be analyzed further.
 Presented approach allows for systematic study of the diffusive particle motion over the  surface with definite  pattern of energies of adatom at lattice sites and energy barriers for jumps  between them.

Below we present the general method of calculations, then 
we show how it works in the case of GaAs(001) surface in two different  $ c(4 \times 4) \beta$ and $ c(4 \times 4) \alpha$ reconstructions of the surface.  We show that  diffusion decreases and becomes more  anisotropic   when $\beta$  reconstruction of the surface  changes to $\alpha$ type.The final expressions for  the diffusion coefficient  allow to identify main diffusion paths in all cases. Effective activation energy value can be calculated for the  diffusion coefficient and for each diffusion  path separately. 

\section{Calculation  of  the  diffusion coefficient.  }
 
To describe random walk of single or uncorrelated adatoms over crystal surface we define the tracer diffusion tensor  \cite{haus,gomer}
\begin{equation}
D_{n,m}= \lim_{t\rightarrow \infty} \frac{1}{4 t}< \Delta r_n(t)  \Delta r_m(t)>
\label{1}
\end{equation}
where $\Delta r_{n[m]}(t)$  is the adsorbate displacement after  time $t$ with respect to
the initial position, along the coordinate  $n[m]=x,y$.  Definition (\ref{1}) 
  in experimental situation corresponds to  diffusion coefficient for the adatom system of low  density.
Adatom diffusion which we want to describe realizes in random walk motion between  different adsorption  sites located
at the crystal surface.
To analyze this motion let us  divide the  lattice of adsorption sites at   crystal surface  into unit cells. 
We assume that each  cell  contains $m$ sites located in positions  $\vec{r}_j^\alpha=\vec{r}_j+\vec{a}_\alpha; \alpha=1,...m$, where $\vec{r}_j$ describes given cell  location and $\vec{a}_\alpha$ is a location of the site  within the cell. Particle at the site described by parameters $(j,\alpha)$ has an  energy $E_\alpha $. The  equilibrium 
 probability of finding adatom at this site is equal  to $P_{eq}(\alpha)=\rho \exp[-\beta E(\alpha)]/\sum_\gamma \exp[-\beta E(\gamma)]$, where sum is over all sites in the unit cell , $\beta=1/(k_B T)$ means inverse temperature parameter and  $\rho$ is density of particles at the surface calculated as number of particles divided by number of unit cells. 

The diffusion motion consists of a series of thermally
activated jumps. The adsorbate in an initial adsorption
state $\alpha$ after a given time 	 escapes to another adsorption
site $\gamma$ with a transition probability per unit time  $W(j,\alpha;l,\gamma)$.
In order to determine the transition probabilities, we apply the transition state theory \cite{TST}, according to which $W(j,\alpha;l, \gamma)$     are functions of the  difference between   energy barrier $\tilde{E}(j,\alpha;l,\gamma)$ for  particle jump between  sites $j,\alpha$ and $l,\gamma$  and $E(\alpha)$,
the particle energy  at the initial  adsorption site
 \begin{equation} \label{W}
W(j,\alpha;l,\gamma)=W_{\alpha,\gamma} =\exp \{-\beta [\tilde{E}(j,\alpha;l,\gamma) -E(\alpha)]\}.
\end{equation} 
Both energies  can be derived from  ab-initio calculations \cite{roehl,2,3}.
Note that  particle  can jump  within one cell or between two different cells $j$ and $l$, so energy $\tilde{E}$ depends on parameters $j, \alpha;l,\gamma$ where $l=j$ when both sites belong to the same elementary cell  or $j\neq l$ when particle changes cell on jumping. $E(\alpha)$ depends only on the type $\alpha$ of the site within given cell. The structure of all possible transitions at the surface can be quite complex, especially when they happen between sites of different energy  located irregularly within the cell. 

Diffusion coefficient defined by (\ref{1}) refers to single particle motion and we assume that our low density system is a collection of independently moving single particles.
The probability that the particle occupies given site   changes with time according to the classical Master equation
\begin{eqnarray}
\label{master}
 \frac{d}{d t} P(j,\alpha;t) =  \sum_{l,\gamma} \Big{[} W_{\gamma,\alpha} P(l,\gamma;t)- W_{\alpha,\gamma} P(j,\alpha;t) \Big{]} 
\end{eqnarray}
where $W_{\gamma,\alpha}$  describe rates of  all possible adatom  jumps over the lattice and fulfill detailed balance condition
\begin{equation}
\label{bal}
W_{\gamma,\alpha}P_{eq}(\gamma)=W_{\alpha,\gamma}P_{eq}(\alpha).
\end{equation}
 All components $W$ can be gathered in one matrix $\hat W$ with elements $-\sum_{l,\gamma} W_{\alpha,\gamma} $ at the diagonal  location  $(j,\alpha;j,\alpha)$ and $W_{ \gamma,\alpha}$  out of the diagonal.

In order to analyze the character of particle  diffusion  it is convenient to make Fourier  transform of the Master equation (\ref{master}) basing on the fact that infinite surface over which the particle is wandering consists of periodically repeated cells. In the wavevector $\vec{ k}$ space the vector describing probability of the site occupation   has components  $P_\alpha (\vec{k};t)=\sum_j \exp(i\vec{k}{\vec{r}_j}^\alpha) P(j,\alpha;t) $. Each component $\alpha$ of this vector  corresponds to one of the sites within unit cell. We have now $m$ equations for  $m$ different components of the occupation  probability vector 
\begin{eqnarray}
\label{master2}
 \frac{d}{d t} P_\alpha(\vec{k};t) =    \sum_{\gamma \neq \alpha}  M(\gamma;\alpha) P_\gamma(\vec{k};t) + M(\alpha;\alpha) P_\alpha(\vec{k};t). 
\label{keq}
\end{eqnarray}
Matrix $\hat M$ is created from   matrix $\hat W$ after  transformation into $k$ space what gives
\begin{eqnarray}
M(\alpha,\alpha)&=&-\sum_{l,\gamma \neq \alpha} W_{\alpha,\gamma} \nonumber\\
M(\gamma,\alpha)&=&W_{\gamma,\alpha} e^{i \vec{k}(\vec{r}_l^\gamma-\vec{r}_j^\alpha)}
\label{m_ab}
\end{eqnarray}
where formulas are written for one, arbitrary  cell  $j$ and summation goes over all sites within this cell and over all neighboring cells $l$.   Note that off-diagonal elements $M(\gamma,\alpha)$ are $\vec{k}$ dependent.

 To solve set of equations (\ref{keq}) we should find $m$ eigenvalues and eigenvectors of matrix $\hat M$. Each eigenvalue describes one dynamical mode responsible for the relaxation of the initial occupation probability towards equilibrium values. The diffusion tensor (\ref{1}) can be obtained  from the one particular eigenvalue
of the transition rate matrix (\ref{m_ab}). It can be
demonstrated   \cite{haus,master,GZ-K}  that for the master equation  (\ref{keq}) there is one and
only one eigenvalue $\lambda_D (\vec{k}) $ such that $\lim_{k \rightarrow 0} \lambda_D(\vec{k}) =0$ and
the real part of all the other eigenvalues is negative. 
This particular  eigenvalue $\lambda_D$ is diffusional eigenvalue, what means that it is  proportional to $\vec{k}^2$ and can be expressed as
\begin{equation}
\label{diff_main}
\lim_{\vec{k} \rightarrow 0}\lambda_D=\lim_{\vec{k}\rightarrow 0} \frac{\vec{w}\hat M \vec{v}}{\vec{w} \vec{v}}=-\vec{k} \hat{D} \vec{k}
\end{equation}
Where $\vec{w}$ is the left eigenvector and $\vec{v}$ the right one.   The  matrix $\hat{M}$ is not hermitian but its elements are related to each other by the rule $M^*(\gamma;\alpha) P_{eq}(\gamma)=M(\alpha;\gamma) P_{eq}(\alpha)$ resulting from  the local equilibrium  balance (\ref{bal}). As a result  eigenvectors of matrix $\hat{M}$  are connected by relation $w^*_\alpha P_{eq}(\alpha) = v_\alpha$. Equation (\ref{diff_main}) is our main formula used  to derive diffusion matrix $\hat{D}$  (\ref{1}).
 On inserting explicit expressions (\ref{m_ab})  into (\ref{diff_main})  and taking into account normalization 
\begin{equation}
\sum_\alpha w_\alpha v_\alpha= \sum_\alpha w_\alpha w^*_\alpha {P_{eq}}(\alpha)= \rho
\label{norm}
\end{equation}
 we can write
\begin{eqnarray}
\vec{k} \hat{D} \vec{k} &=&\frac{1}{\rho}\lim_{\vec{k} \rightarrow 0} \sum_{\alpha,\gamma} w_\gamma M(\alpha,\gamma) v_\alpha  
\\& =&\frac{1}{\rho} \lim_{\vec{k} \rightarrow 0} \sum_{\alpha,\gamma} (w_\alpha-e^{\vec{k}(\vec{r}_l^\gamma-\vec{r}_j^\alpha)} w_\gamma )W_{\alpha,\gamma} v_\alpha \nonumber.
\end{eqnarray}
And when  consequences of detailed balance for $M({\alpha},{\gamma})$ and eigenvectors $\vec{w}, \vec{v}$ are  taken into account we have
\begin{eqnarray}
&\vec{k} \hat{D}\vec{k}& = \frac{1}{\rho} \lim_{\vec{k}  \rightarrow 0} \big{[} \sum_{\alpha>\gamma} (w_\alpha-e^{\vec{k}(\vec{r}_l^\gamma-\vec{r}_j^\alpha)} w_\gamma )W_{\alpha,\gamma} P_{eq}(\alpha) w^*_\alpha   \nonumber \\
&+&( e^{\vec{k}(\vec{r}_l^\gamma-\vec{r}_j^\alpha)} w_\gamma- w_\alpha )W_{\alpha,\gamma} P_{eq}(\alpha) w^*_\gamma  e^{-\vec{k}(\vec{r}_l^\gamma-\vec{r}_j^\alpha)} \big{]} \nonumber \\
&=&\frac{1}{\rho}\lim_{\vec{k}  \rightarrow 0} \sum_{\alpha>\gamma} |w_\alpha-e^{\vec{k}(\vec{r}_l^\gamma-\vec{r}_j^\alpha)} w_\gamma |^2 W_{\alpha,\gamma} P_{eq}(\alpha) 
\label{last}
\end{eqnarray}
Our diffusional  eigenvalue $\lambda_D$  in the limit of small $\vec{k}$ is proportional to $\vec{k}^2$ and  has the  lowest   absolute value of all eigenvalues. It can be found  by  the variational method on assuming properly parameterized eigenvector. We choose 
\begin{eqnarray}
\label{vec}
v_\alpha&= &P_{eq} (\alpha) e^{i \vec{k} \vec{\phi}_\alpha} \nonumber \\
w_\alpha&=& e^{-i \vec{k} \vec{\phi}_\alpha}
\end{eqnarray}
The above form of  eigenvectors provides proper normalization and introduces variational  parameters $\vec{\phi}_\alpha$ that are  coupled to wave-vector $\vec{k}$ so they can influence diffusion coefficient. Similar choice of variational parameters was shown to be  a good one  in description of collective diffusion  at non-homogeneous surfaces in Refs. \onlinecite{Z-KG1,BZ-KG1,Z-KBG,Z-KG2}.

On using explicit expression for eigenvectors  (\ref{vec}) in the last equation (\ref{last})  we obtain  variational  formula for diffusion coefficient $\hat{ D}_{var}$ 
\begin{eqnarray}
\label{final}
&\vec{k} \hat {D}_{var} \vec{k}& =\lim_{\vec{k} \rightarrow 0} \sum_{\alpha>\gamma} W(\alpha;\gamma) P_{eq}(\alpha) [e^{i \vec{k} \phi_\alpha}-e^{\vec{k}(\vec{r}_l^\gamma-\vec{r}_j^\alpha) +\phi_\gamma)}]^2 \nonumber \\
&=   &\sum_{\alpha>\gamma} W(\alpha;\gamma) P_{eq}(\alpha)[\vec{k} ( \vec{\phi}_\gamma+\vec{r}_l^\gamma-\vec{r}_j^\alpha-\vec{\phi}_\alpha)]^2
\end{eqnarray}
Now, for the final expression for $\hat{D}_{var}$ we should minimize the above equation with respect to all independent parameters $\vec{\phi}_\alpha$. Each  separate type of the surface site $\alpha$ has its individual vector of  phases, so all together we have two times more phases than  sites in the elementary cage. However one of vector parameters $\vec{\phi}$ can be set to be $\vec{0}$, because as it can be seen in  (\ref{final})  only phase differences  count, what means that we can move all phases by the same quantity without changing the result. Moreover  phases for some sites are identical  due to the system  symmetry. In such a way the number of parameters can be reduced and as a result the problem is simplified. In the following sections we will show how the procedure works in practice on studying diffusion of  Ga adatoms over GaAs(001)  c(4$\times$4) surface.
\begin{figure}
\includegraphics[width=6cm, height=6cm]{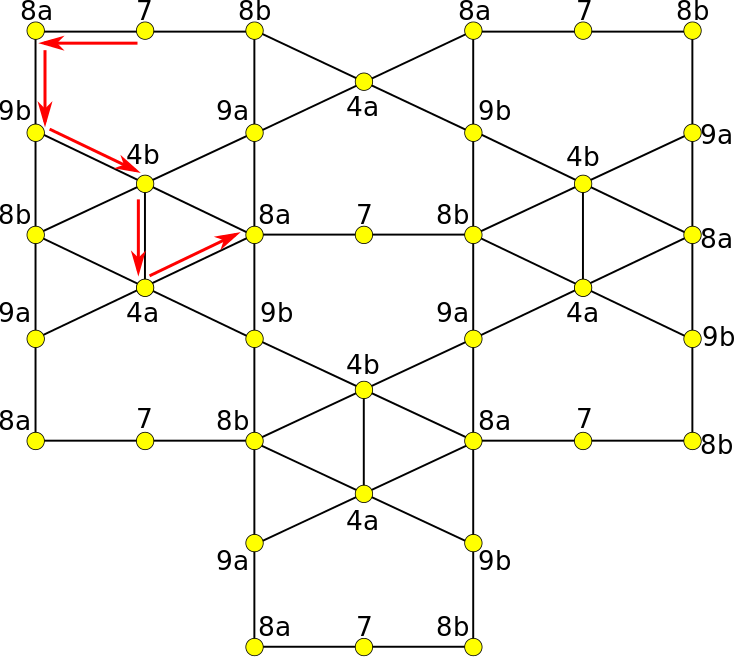}
\caption{Binding sites of Ga atoms at the GaAs(001)-c($4\times4$) $\beta$ surface.\label{lattice}}
\end{figure}
\begin{figure}
\includegraphics[width=6cm]{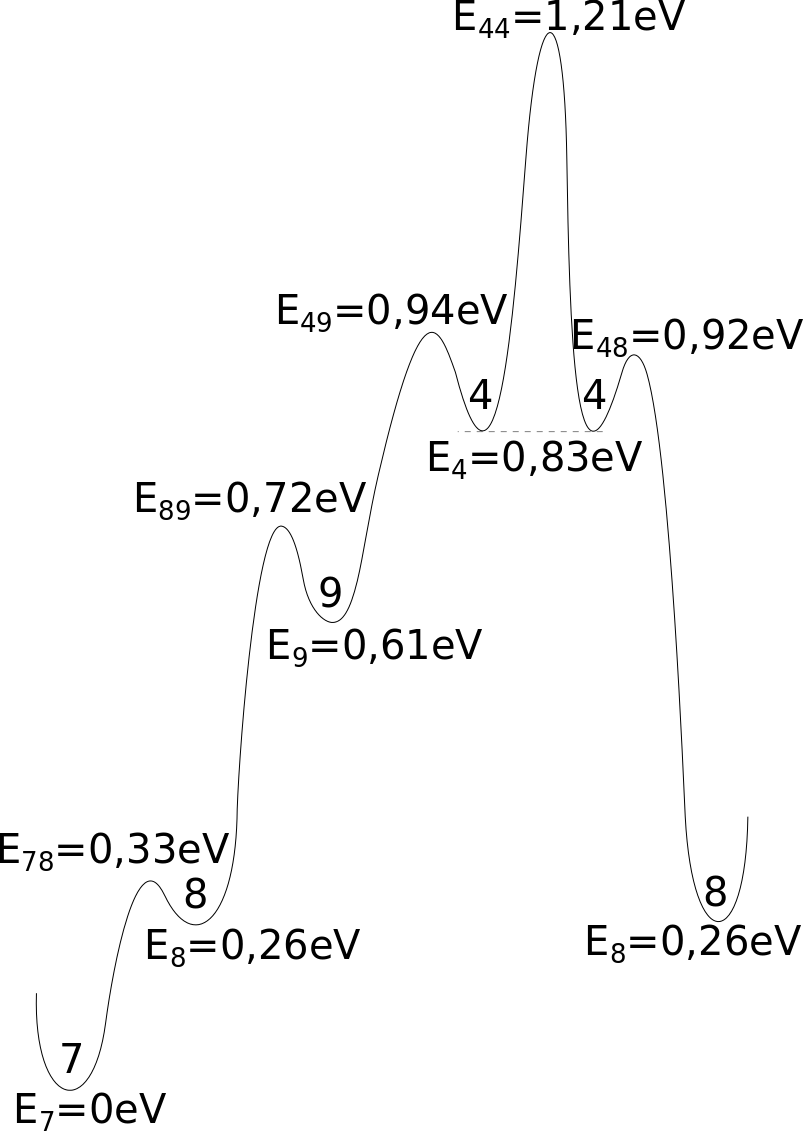}
\caption{Energy landscape cross-section along the marked path in Fig. \ref{lattice}. It  contains all the possible Ga adatom jumps at the surface.\label{landscape}}
\end{figure}
\section{Diffusion coefficient for $\rm Ga$ adatom at $\rm GaAs(001)-c(4 \times 4) \beta$ reconstruction}
Systematic study of energies of  Ga atom  at binding  sites at $ \rm GaAs(001)-c(4 \times 4)\beta$  surface reconstruction and of barriers for diffusion between them have been first shown  in Ref \onlinecite{2}. Lately more detailed map of  the surface energy was  presented in  Ref. \onlinecite{roehl}. We used  the energy pattern calculated in this last  article as a basis to the analysis of the diffusion behavior of adsorbed  Ga atom.  

 The scheme of the  lattice for Ga adsorption sites at  $\rm GaAs(001) - c(4 \times 4)\beta$ surface  is shown  in Fig.\ref{lattice} with numeration from  Ref.\onlinecite{roehl}. When Ga atom was relaxed from 3$\AA$ above the surface  seven different types of minimum energy sites were found.  The lowest one with number   $7$ (described also as $6$ \cite{roehl})   lies in the trench  between As dimers at the reconstructed surface and so are four other sites:  $8a,8b$ and $9a,9b$.  As it was shown in Ref. \onlinecite{roehl} position and depth  of the adsorption sites at hills of As dimers depends on the height from which   Ga atom is relaxed.
 However if we stay at height  of 3$\AA$ the surface is not deformed and there is only one additional adsorption site at the surface in two versions $4a$ and $4b$. With the same numbers and different letters we marked sites of  the same depth and the same jump rates out of the site, however  with  different  space orientation, rotated or reflected  like in the case of $8a$ and $8b$ (see Fig.\ref{lattice}). 
All energies that decide about jump rates between our sites can be presented along path marked in Fig.\ref{lattice}
  and energy landscape along this path is  shown in Fig.\ref{landscape}. It can be seen that the site of the lowest energy is $7$ and there are   four different adsorption   site energy values at this surface. All together  there are seven sites - one of energy$ E_7$, and two types of energies $E_4, E_8$ and $E_9$ in one elementary cage. 

Calculated energy barriers determine the probability of each single particle jump. An analysis of  barrier heights allows to find the easiest jump path, what have been done in Ref. \onlinecite{roehl}. However when we look for the path with minimal energy barriers it does not necessarily reflect possible diffusional behavior of single particle. When particle moves randomly in  the potential landscape like this in Figs. \ref{lattice}, \ref{landscape} it jumps forward and backward with frequency proportional to the jump rate. As it can be seen  when particle jumps down,  into the site of lower  energy,  its return is more  difficult due to high energy barrier. It means that not only transition rates of jumps but also site occupation probabilities at equilibrium should be also important in the long distance diffusive motion. Our formula (\ref{final}) contains all transition rates as well as equilibrium occupation probabilities for all possible sites. Diffusion coefficient matrix  describes long distance particle diffusion at equilibrium conditions. Below we calculate tensor of diffusion coefficients  for  single particle motion on using  the variational approach. Then we  compare diffusion in different directions and finally identify dominant diffusion paths. As we will see only some of them  agree with the  paths of minimal energy barriers. 

In Fig.\ref{lattice} we plot all transitions between  sites that are taken into account in the calculations below. Energy barriers for these transitions were taken from Ref.\onlinecite{roehl}. There is  one more  possible transition path not shown in Fig.\ref{lattice}  but calculated  in Ref.\onlinecite{roehl}. It is path directly from  site $7$  through low lying site $5$ to the site $4$ with the activation energy as high as $1.10 eV$. In comparison with other energy barriers for jumps  in the  systems this transition rate can be neglected at experimentally   achieved temperatures. As a result there is also no need to  consider any additional sites, like $5$, hence we end up with four different energies of adsorption sites.

 According to the rules of  statistical mechanics equilibrium probability at given site is
\begin{eqnarray}
P_{eq}^{\alpha}=\frac{\rho e^{-\beta E_\alpha}}{2e^{-\beta E_4}+e^{-\beta E_7}+2e^{-\beta E_8}+2e^{-\beta E_9}}
\label{peq}
\end{eqnarray}
when the site occupation is normalized within elementary cage $ \sum_{\alpha}P_{eq}^\alpha=\rho$.  
\begin{figure}
\begin{center}
\includegraphics[width=6cm]{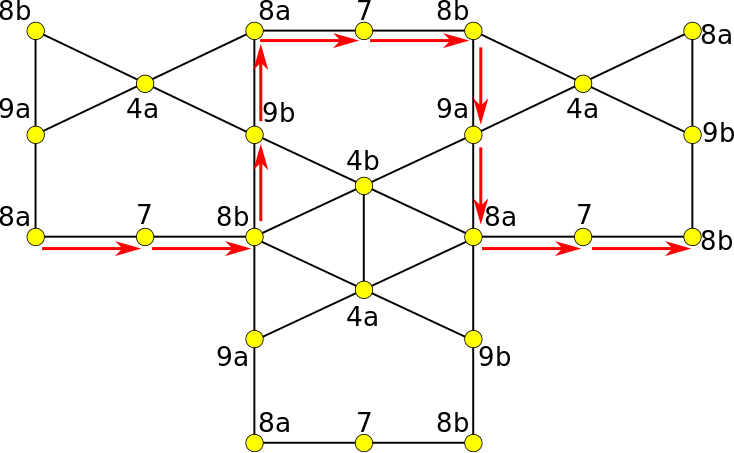}
\includegraphics[width=6cm]{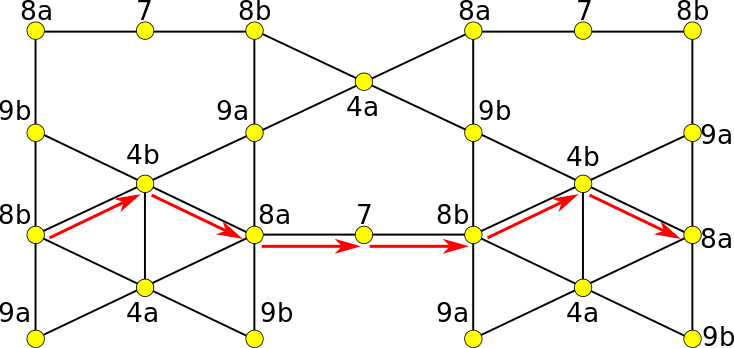}
\includegraphics[width=6cm]{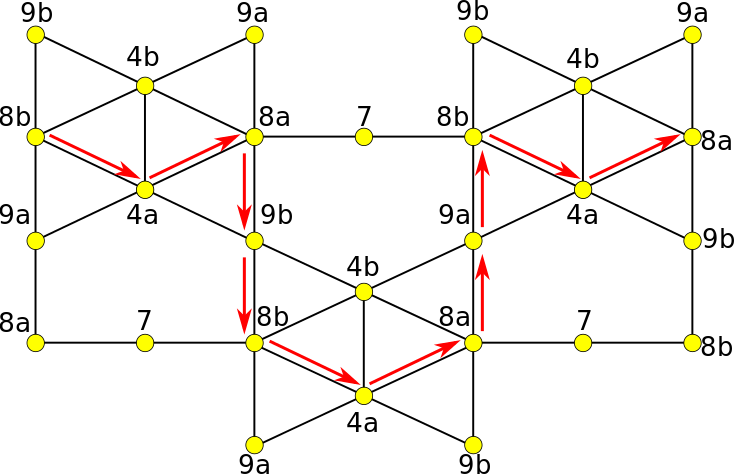}
\includegraphics[width=6cm]{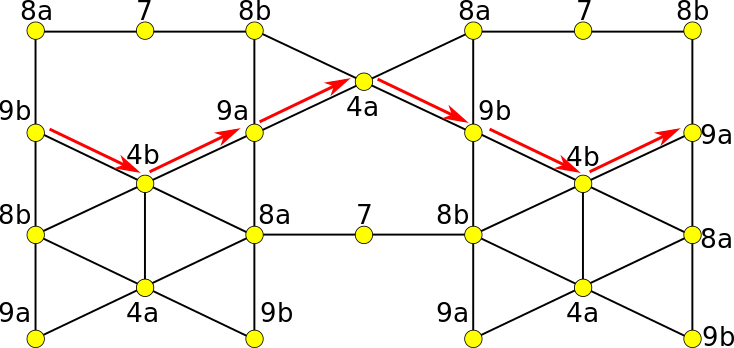}
\end{center}
\caption{Diffusion Ga adatom paths  along the x-direction at GaAs(001)-c($4\times4$) $\beta$ surface}
\label{paths_x}
\end{figure}
\begin{figure}
\begin{center}
\includegraphics[width=4.5cm]{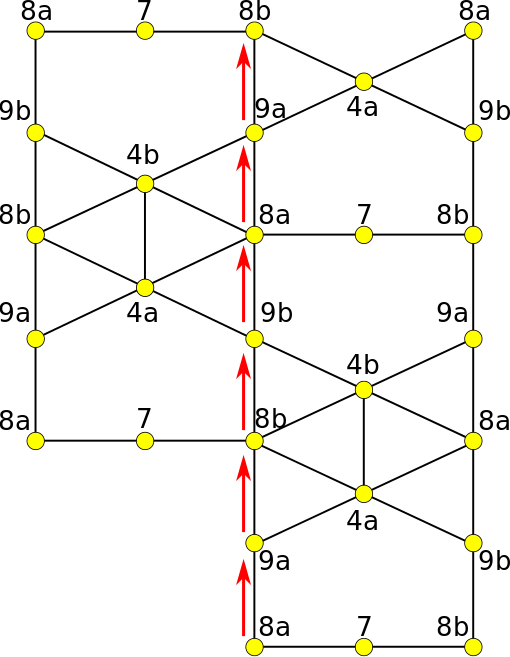}
\includegraphics[width=4.5cm]{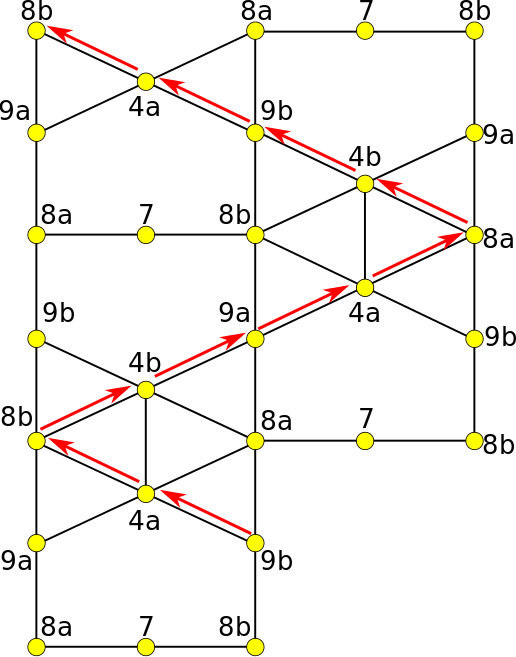}
\includegraphics[width=4.5cm]{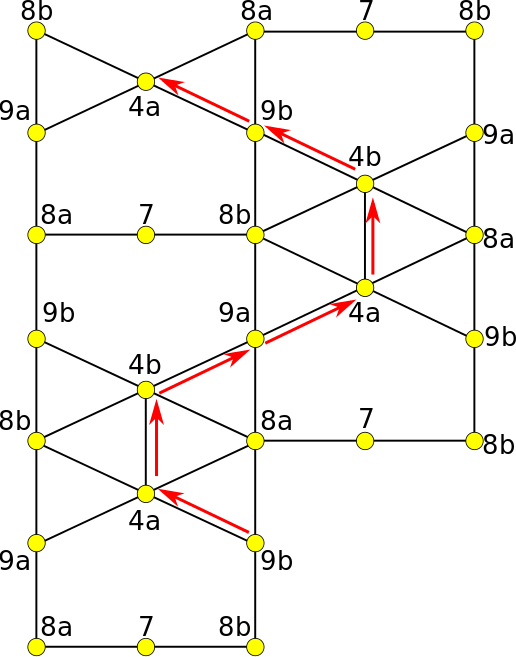}
\end{center}
\caption{Diffusion Ga adatom paths along the y-direction at GaAs(001)-c($4\times4$)$\beta$ surface}
\label{paths_y}
\end{figure}
Equilibrium probabilities  (\ref{peq}) should be then put   into the  variational formula (\ref{final}) for  diffusion matrix. In general there are two variational parameters $\phi_x^\alpha$ and $\phi_y^\alpha$ per each site type - one coupled to the direction $k_x$ and the other to $k_y$. 
The lattice has reflection symmetry with respect to  the directions $x$ and $y$ and according to this symmetry $\phi_{x,y}^{9a}=-\phi_{x,y}^{9b}$ - closest bonds are inverted in both directions, whereas $\phi_x^{8a}=-\phi_x^{8b}$, $\phi_y^{8a}=\phi_y^{8b}$ and inversely for the  site $4$, $\phi_x^{4a}=\phi_x^{4b}$,$\phi_y^{4a}=-\phi_y^{4b}$. As mentioned before phases at one of sites can be set freely, so the number of parameters is reduced to six. Axes $x$ and $y$ are  the main directions of the diffusional tensor $\hat{D}$ for the lattice with the  described symmetry, hence only diagonal parameters $D_{xx}=D_x$ and $D_{yy}=D_y$ are nonzero and  each value at diagonal depends only on  parameters coupled to the corresponding direction. Finally the following values of diffusion coefficients were found
\begin{eqnarray}
D_{x}^{\beta}&=&2 \Big{[}\frac{(W_{87}W_{89}+4W_{84}W_{87}+2W_{84}W_{89})P_{eq}^{8}}{2W_{84}+W_{87}+2W_{89}} \nonumber \\&+&W_{49}P_{eq}^{4}\Big{]} {a^2}
\label{Dx}
\end{eqnarray}
\begin{equation}
D_{y}^{\beta}=2\Big{[}W_{89}P_{eq}^{8}+\frac{W_{49}(W_{44}+W_{48})P_{eq}^{4}}{W_{44}+W_{48}+W_{49}}\Big{]} a^2
\label{Dy} 
\end{equation}
where $W_{\alpha,\beta}$ is given by Eq. (\ref{W}) and appropriate energies can be found in Fig. \ref{landscape}. Lattice unitary length $a=5.6\AA$.

In the above expressions each of  terms can be understood as a contribution of a certain path to the total diffusion. 
Let us  note that among all paths only one in  direction $x$ (between sites 4 and 9) and one in  direction $y$ (between sites 8 and 9) are independent from other ways of diffusion. Remaining transition rates of Eq. (\ref{Dx}) and Eq.(\ref{Dy}) are all connected in one component. The  numerator  of both these expression is a sum of separate   terms. For example in Eq. (\ref{Dx}) we can write first component as a sum of three  terms, each of them containing two different rates in the numerator and the same rates   in the denominator plus additional third one. Thus we can say that each of these terms represents path which goes through the main two links and it is slightly modified by the presence of other ways of diffusion. We can now plot all possible ways of diffusion.
There are four paths in  x-direction and three paths in  y-direction. We show them in Figs \ref{paths_x} and \ref{paths_y}.

\begin{figure}
\begin{center}
\includegraphics[angle=0,width=7cm]{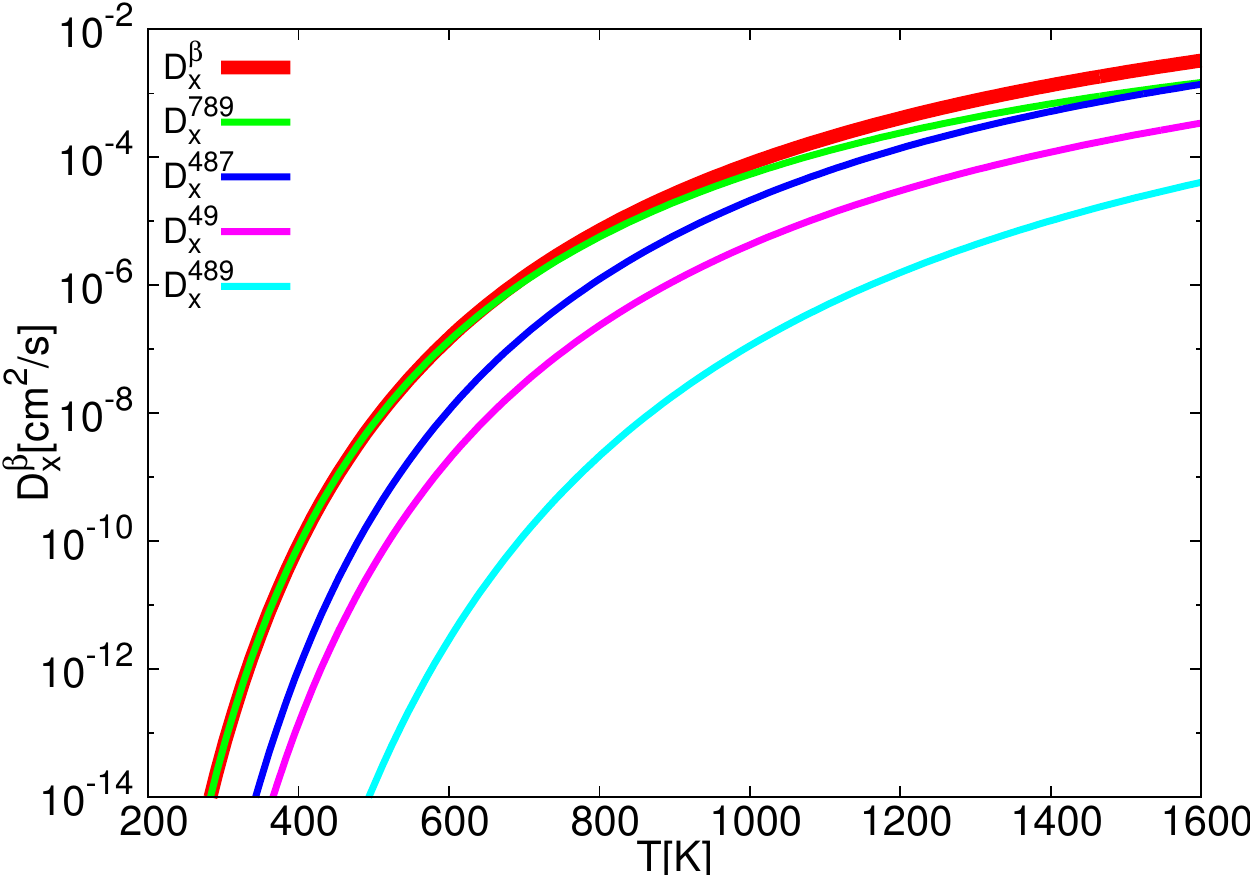}
\includegraphics[angle=0,width=7cm]{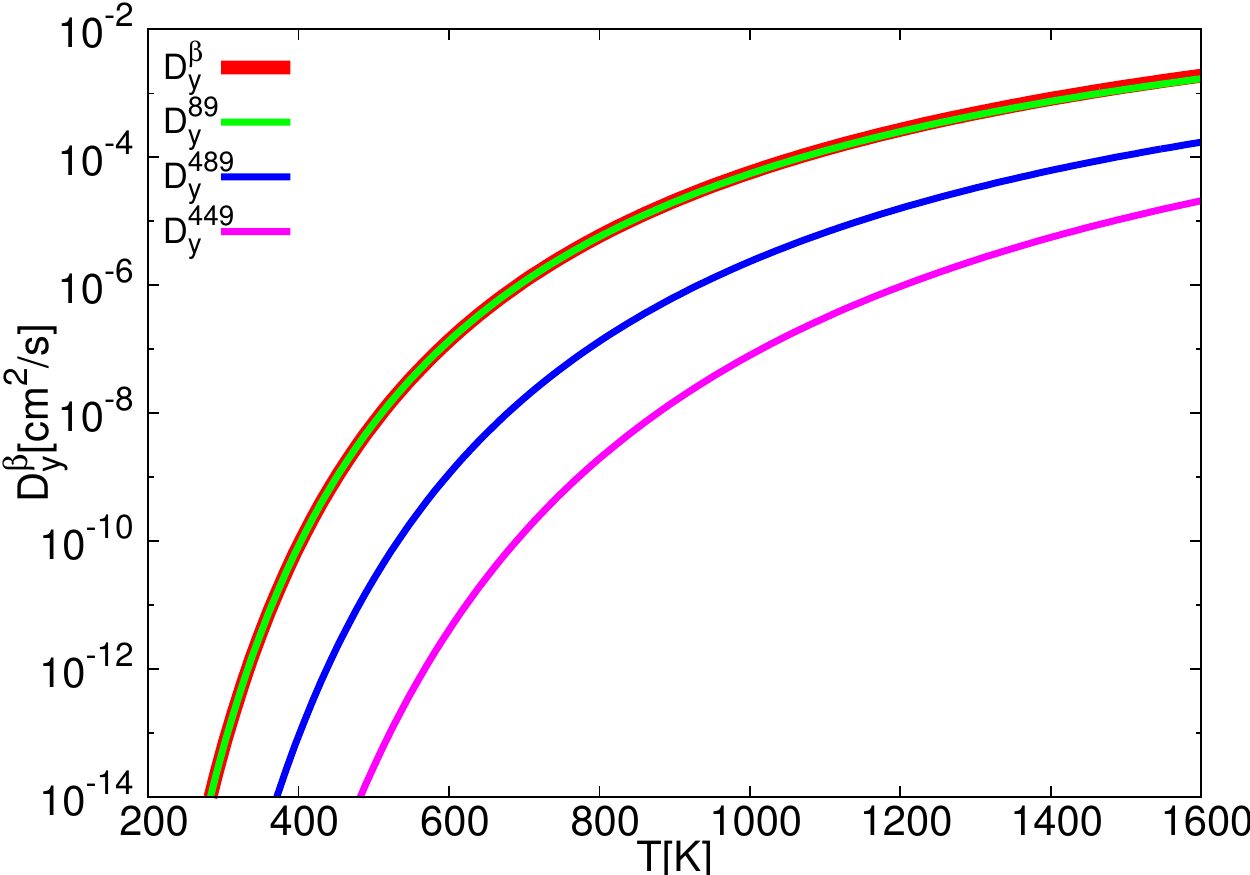}
\end{center}
\caption{Dependence of the diffusion coefficients $D_x$ and $D_y$ on the temperature for Ga atom at GaAs(001)-c($4\times4$) $\beta$ surface\label{diff_plot}}
\label{diff}
\end{figure}
\begin{figure}
\begin{center}
\includegraphics[angle=0,width=7cm]{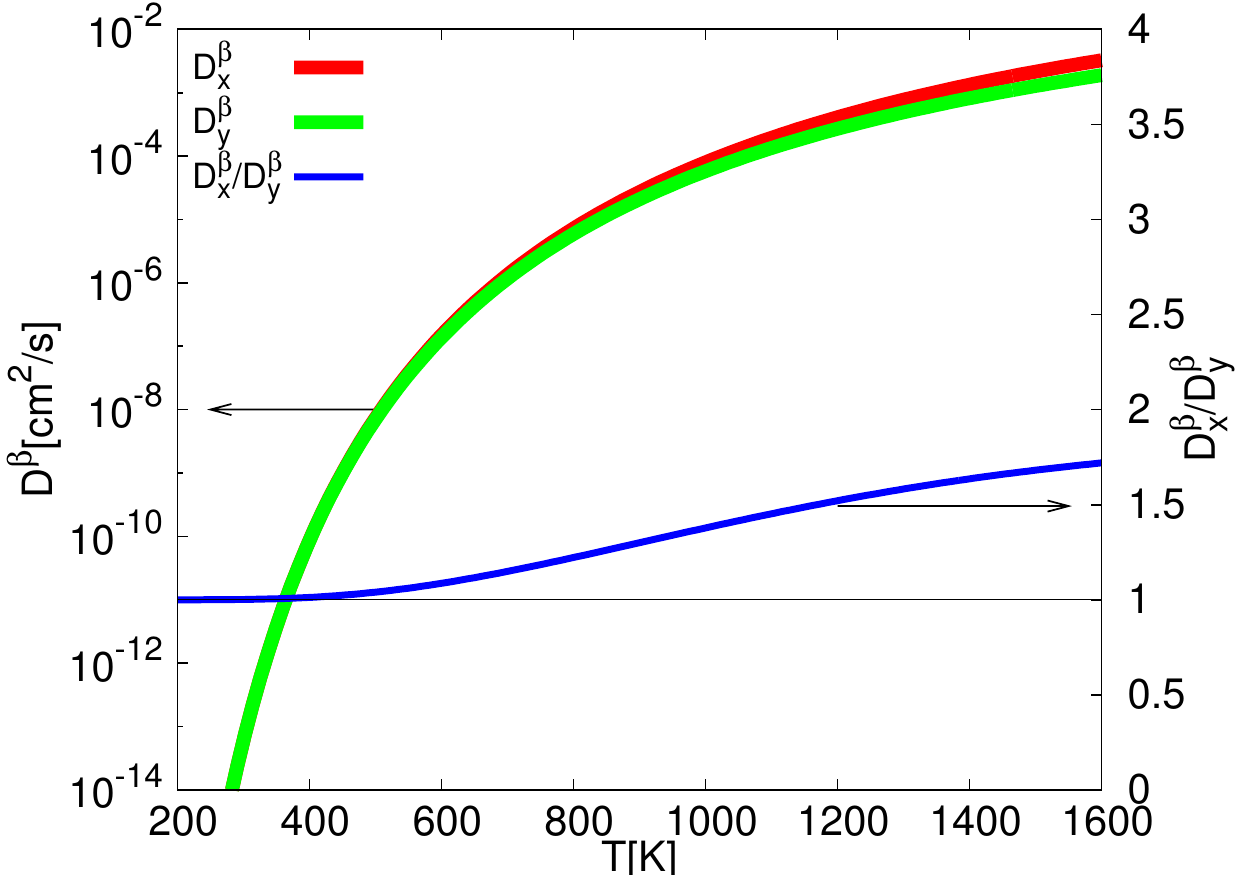}
\caption{Comparison of the diffusion coefficients $D_{x}^{\beta}$ and $D_{y}^{\beta}$  as  a temperature function in logarithmic scale. The scale of the anisotropy coefficient  is shown on the right.
\label{compare}}
\end{center}
\end{figure}
\begin{figure}
\begin{center}
\includegraphics[angle=0,width=7cm]{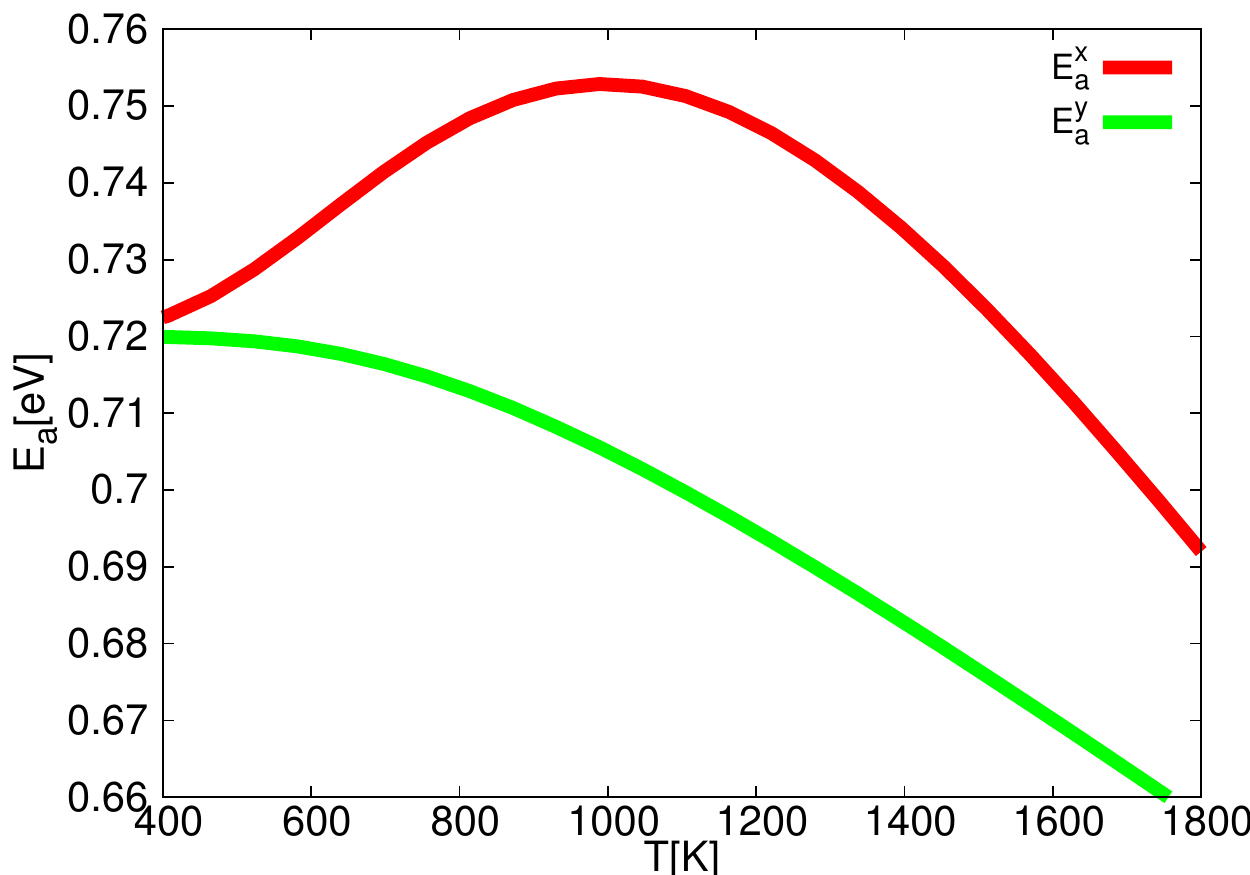}
\end{center}
\caption{Temperature dependence of the activation energies calculated with use of (\ref{ea}) for total diffusion coefficients $D^\beta$ in both directions. \label{activation_fig}}
\end{figure}
At first let us analyze diffusion along direction x. The first plotted path contains only $W_{87}$ and $W_{89}$ transition rates and it is the same path as was identified as the one with the lowest activation energies for jumps in Ref.~\onlinecite{roehl}. In Fig.\ref{diff_plot} we plotted temperature dependence  of total diffusion coefficient in both directions and contribution to this value of  each identified path.  Our results confirm that path found   in Ref.~\onlinecite{roehl} is the most important for large  temperature range. The next important path for the diffusion coefficient is the one that goes through the hill site~4. Diffusion along this path becomes larger  than the first one at higher temperatures.  As explained above these paths   are not entirely independent, they are linked by denominator in one expression. This expression contains also third path, which has the lowest diffusion values. The independent path (49) is   not the fastest one, but it is interesting because it bypasses  site 7, the one of the lowest energy, 
what means that once  particle comes out of the lowest energy site it can slide over states with  higher energy.
Such  independent  path and  avoiding site~7 is the most important one in direction y. It appears that due to the symmetry of the lattice all possible paths along direction y bypass the lowest site as can be seen in Fig.~\ref{paths_y}. However according to plots in  Fig.~\ref{diff_plot} only one path has significant effect on the overall diffusion in  direction y.

It is interesting to compare diffusion along x and y direction. Components in both directions are plotted  in Fig \ref{compare}. It can be seen that  diffusion in x direction is  faster than that in direction y. Note that  this difference seems to be small only due to the logarithmic scale used in this figure. At the same figure ratio between  $D_x$ and $D_y$ is plotted in linear scale. For low temperatures ratio between both diffusion components is close to one, i.e. in figure it is close to thin  horizontal line  of height 1. This agrees with Monte Carlo simulations in Ref.~\onlinecite{2} where diffusion at temperature 470K was found to be  isotropic. However at higher  temperatures  diffusion becomes an anisotropic process with $D_{x}$ almost two times higher than $ D_{y}$. 

When we ignore in  (\ref{Dx}) and (\ref{Dy}) all contributions  from sites located at dipole structure we have expressions
\begin{eqnarray}
D_x=2 \frac{W_{87}W_{89}P_{eq}^8}{W_{87}+2W_{89}}a^2  \nonumber \\
D_y=2 W_{89} P_{eq}^8 {a^2}
\end{eqnarray}
which can be explicitly used to reproduce Monte Carlo (MC) data from (\onlinecite{2}). When we put into the above formulas exact energy barriers that were used in  Ref. \onlinecite{2} we get  $D_x=1.857^. 10^{-8}$  and  $D_y=1.859^.10^{-8}$ comparing with MC results $1.74^.10^{-8}$ in  direction x and  $1.66^. 10^{-8}$ in direction y. 

In order to analyze diffusion process more precisely we use parametrization
\begin{eqnarray}
D_{x}&=&\nu_{x} e^{-\beta {E_A}^{x}} \nonumber \\
D_{y}&=&\nu_{y} e^{-\beta {E_A}^{y}}
\end{eqnarray}
where both activation energy $E_A$ and  prefactor $\nu$  can weakly   depend on temperature.
We calculate activation energies on  using the formula:
\begin{equation}
\label{ea}
E_{A}=-\frac{\partial\ln D}{\partial\beta}
\end{equation}
In Fig \ref{activation_fig} we plot   temperature dependence of activation energies for both coefficients $D_{x}$, $D_{y}$. 
It can be seen that activation energy does not change much within the range of temperature which is important in the experiment . In both directions it starts with the same  value $0.72eV$  
then $E_A^x$ slightly goes up has its maximum at  $1000K$ and then   decreases, whereas    $E_A^y$  goes down within presented temperature range. It is interesting that even if $D_y$ is lower than $D_x$, the effective activation energy responsible for this direction is also lower that $E_A^x$. It means that this is not activation energy but prefactor $\nu$ that  decides about the observed anisotropy.
Looking at  curves in Fig.~\ref{diff} it is easy to understand that diffusion in the direction $x$ is dictated by (789) direction at lower temperatures and then it is more and more dependent on the second channel (487), whereas diffusion along $y$ direction is given almost totally by one, dominant component (89). The number of important diffusion paths contributes to prefactor, thus compensating the effect from the activation barrier difference.

We can see that at the surface with several adsorption sites and a complicated lattice of transitions  few most important paths can be identified. In close to equilibrium conditions particles diffuse along these paths.  We have shown that activation energy that characterizes given diffusion path is not a simple sum of energy barriers for individual jumps between successive adsorption sites. Equilibrium diffusion coefficient strongly depends on the occupation  probabilities that always accompany corresponding  transition rates. However the slowest jump in a row is the  one that decides about the value of the diffusion coefficient. Such property can be seen in the  shape of  our formulas, where sums of the   reciprocals of successive  rates are present. Moreover diffusion process not necessarily  happens through the lowest adsorption site, like our (89) path for diffusion along $y$ direction.

\begin{figure}
\includegraphics[width=7cm]{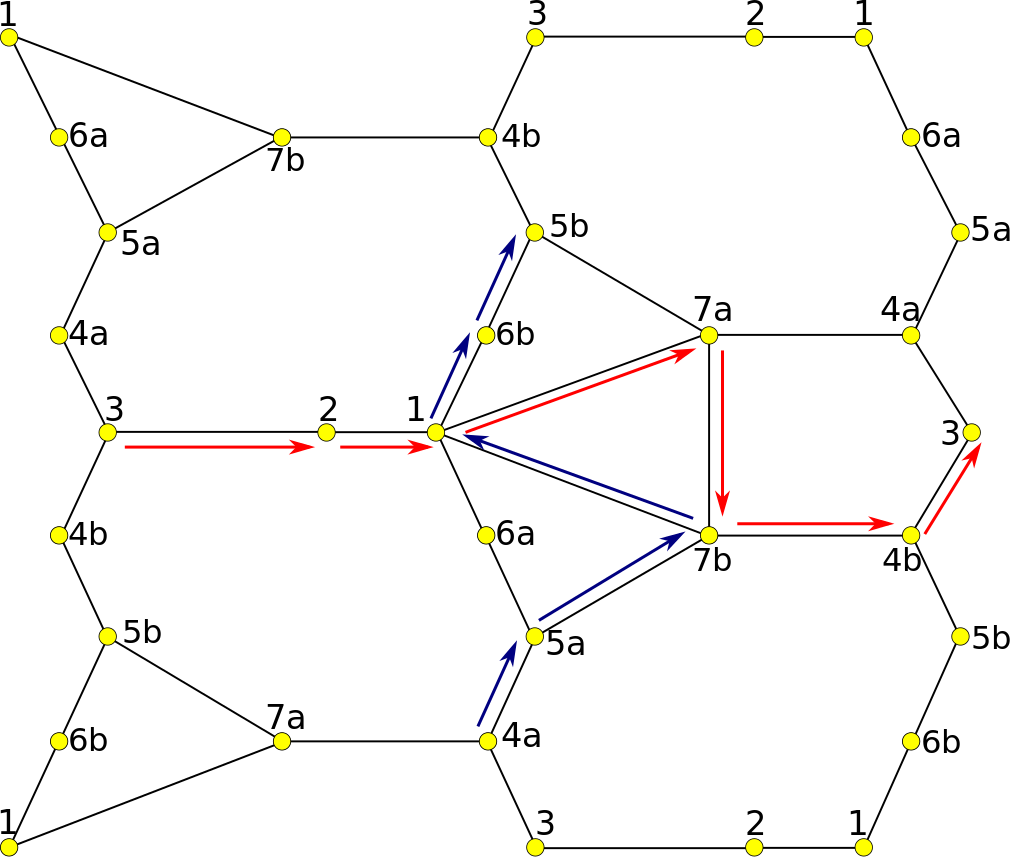}
\caption{Binding sites of Ga atoms at GaAs(001)-c($4\times4$) $\alpha$ surface.\label{lattice2}}
\end{figure}\begin{figure}
\includegraphics[width=6cm]{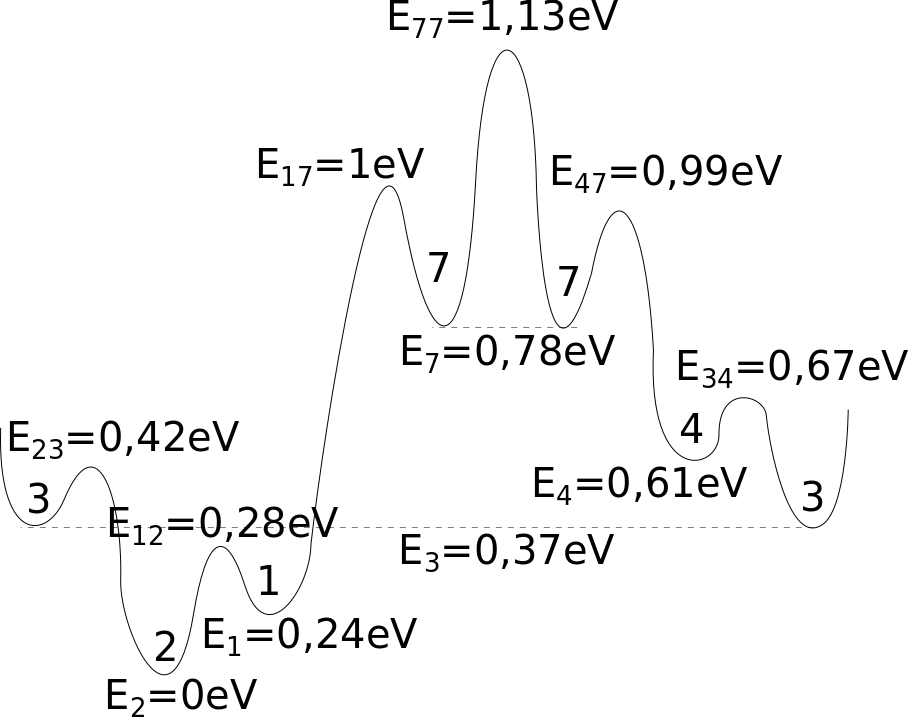}
\includegraphics[width=6cm]{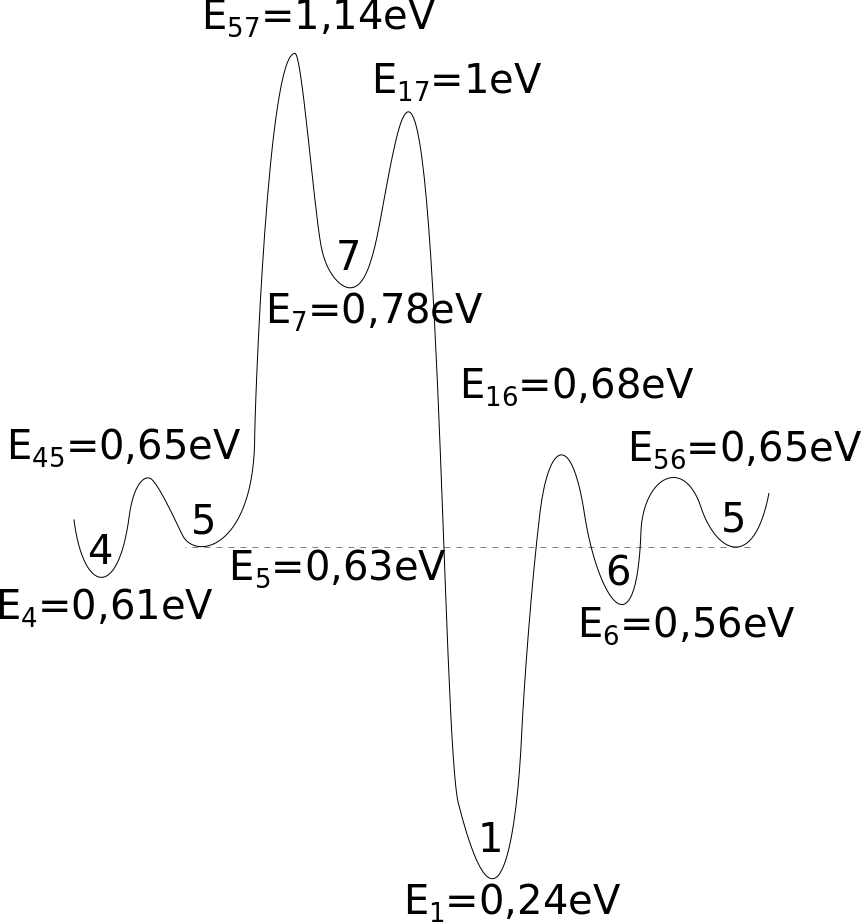}
\caption{Energy landscape along the marked paths in Fig.~\ref{lattice2}. They contain all the possible jumps of Ga adatom at the surface.\label{landscape2}}
\end{figure}
\begin{figure}
\includegraphics[width=6cm]{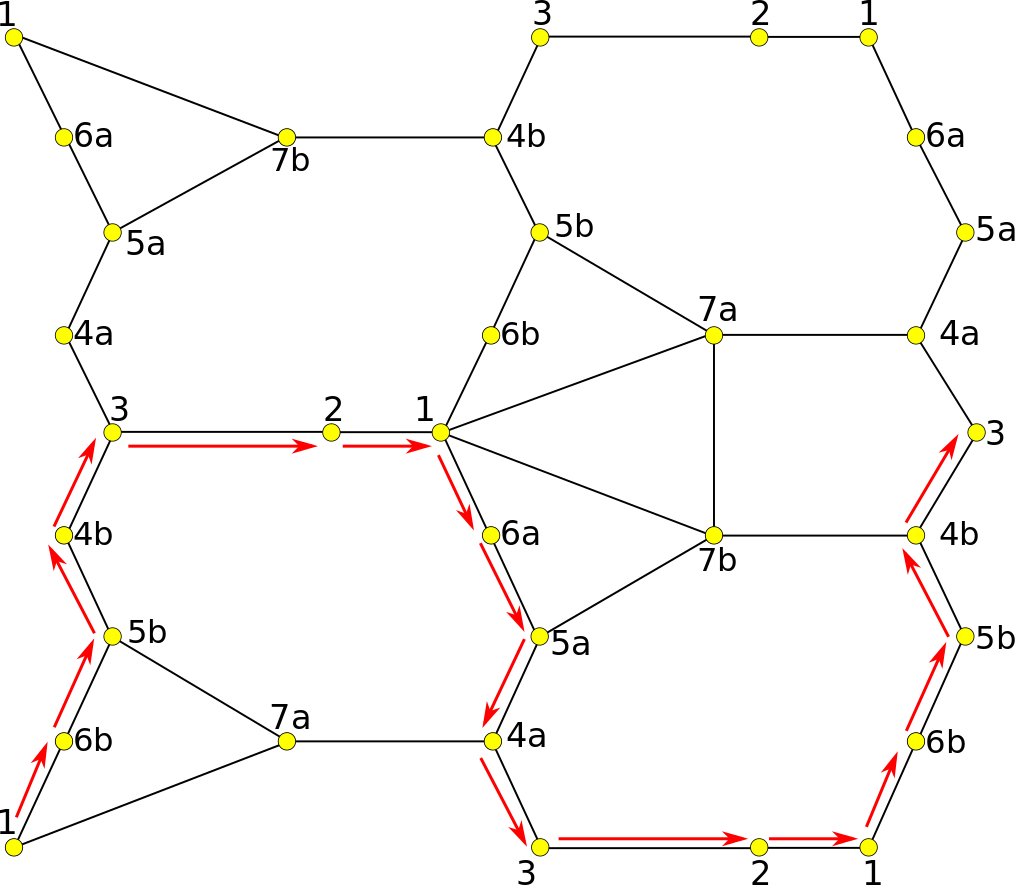}
\includegraphics[width=6cm]{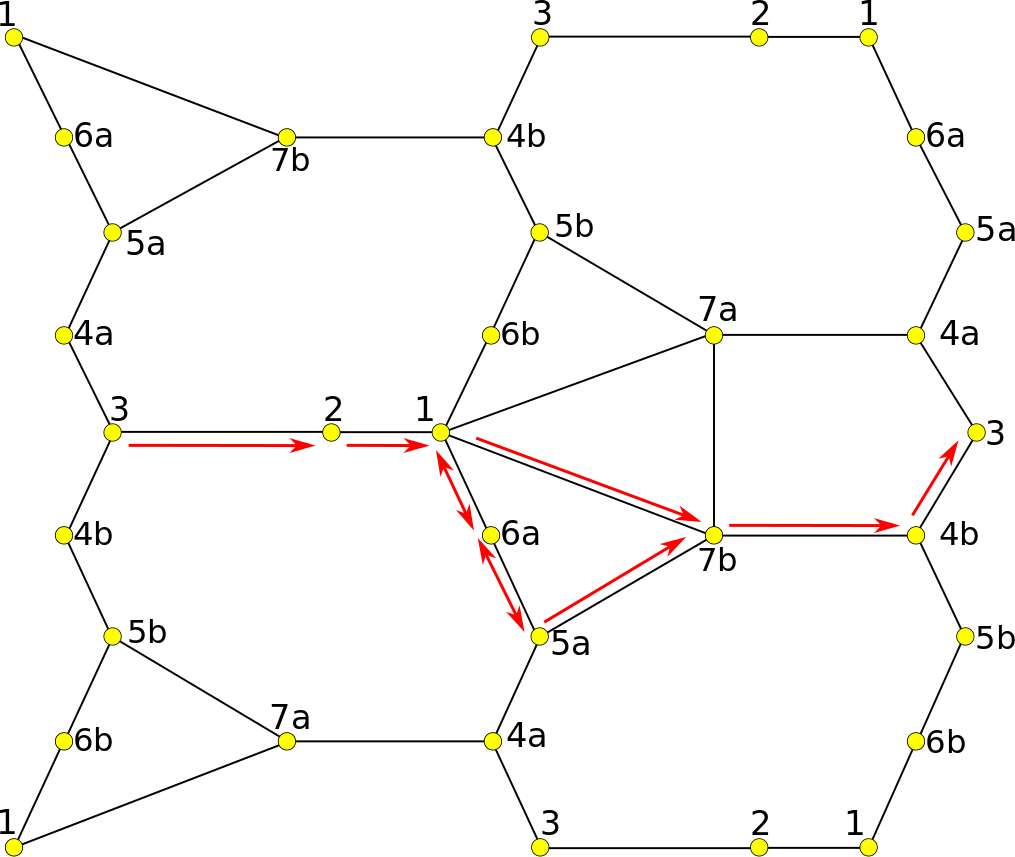}
\includegraphics[width=6cm]{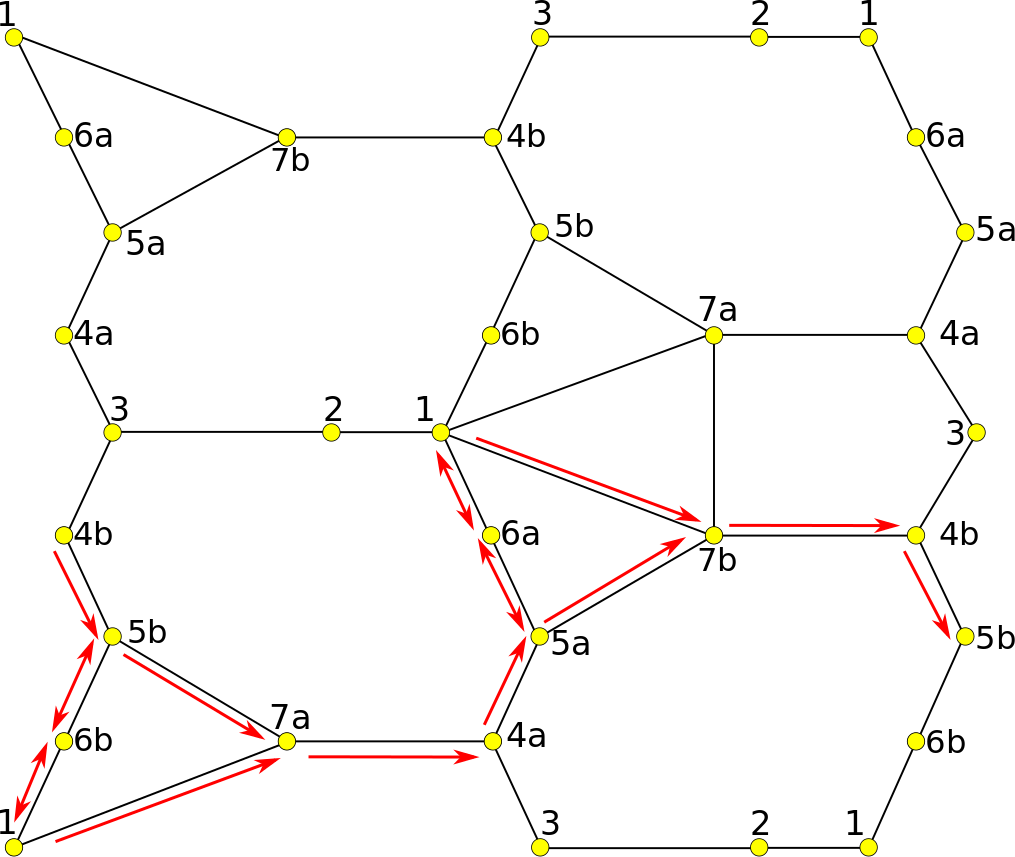}
\caption{Different Ga adatom diffusion paths in  direction x over GaAs(001) $(4\times 4) \alpha$ reconstructed surface.\label{path2x}}
\end{figure}
\begin{figure}
\includegraphics[width=6cm]{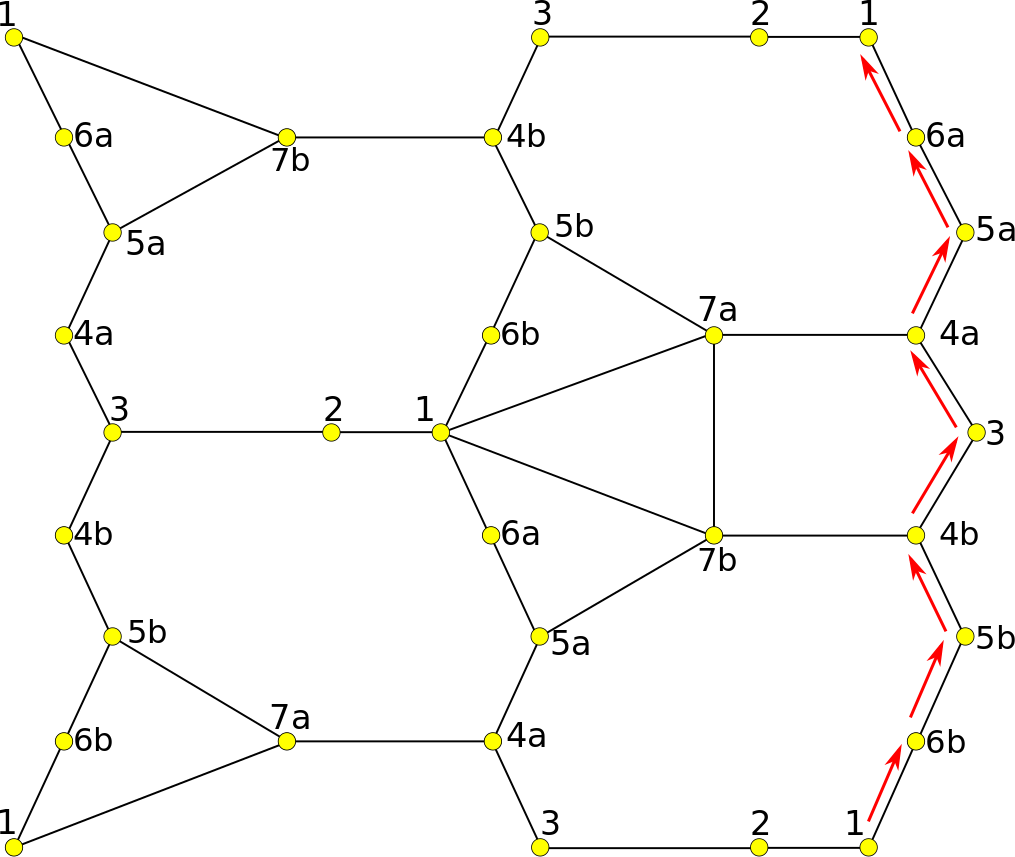}
\includegraphics[width=6cm]{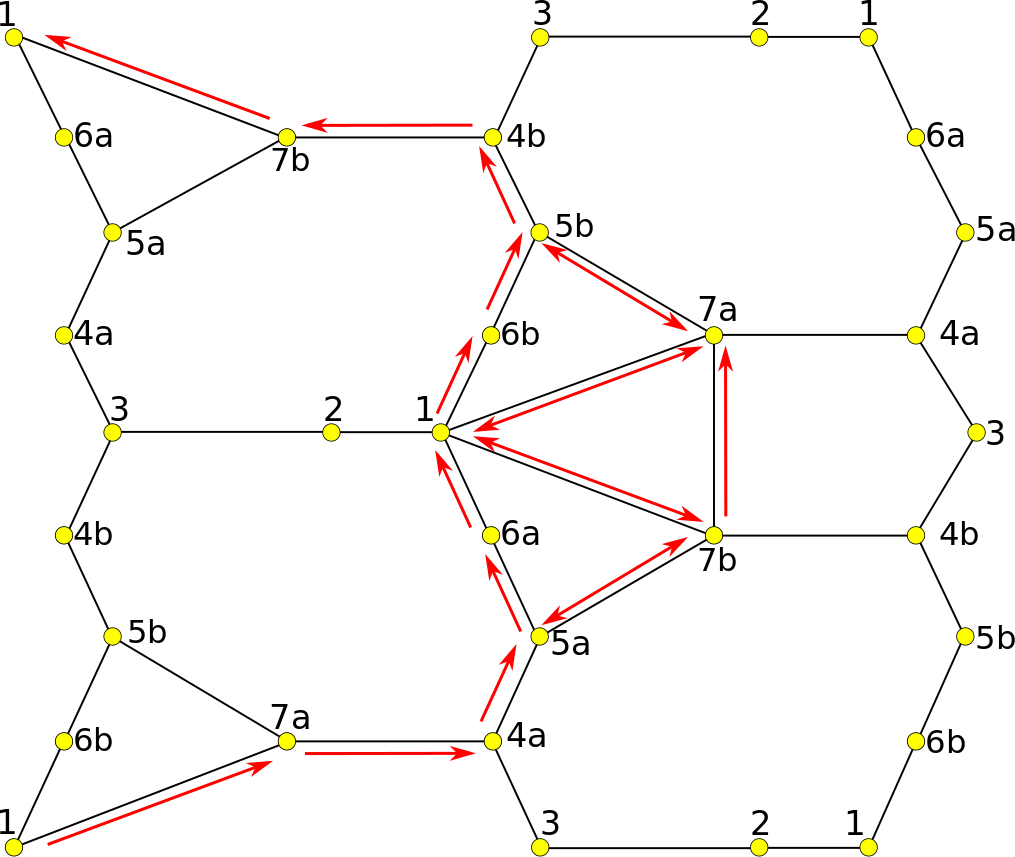}
\caption{Possible Ga adatom  diffusion paths in direction y over GaAs(001) $(4\times 4) \alpha$ reconstructed surface.\label{path2y}}
\end{figure}
\begin{figure}
\includegraphics[width=7cm]{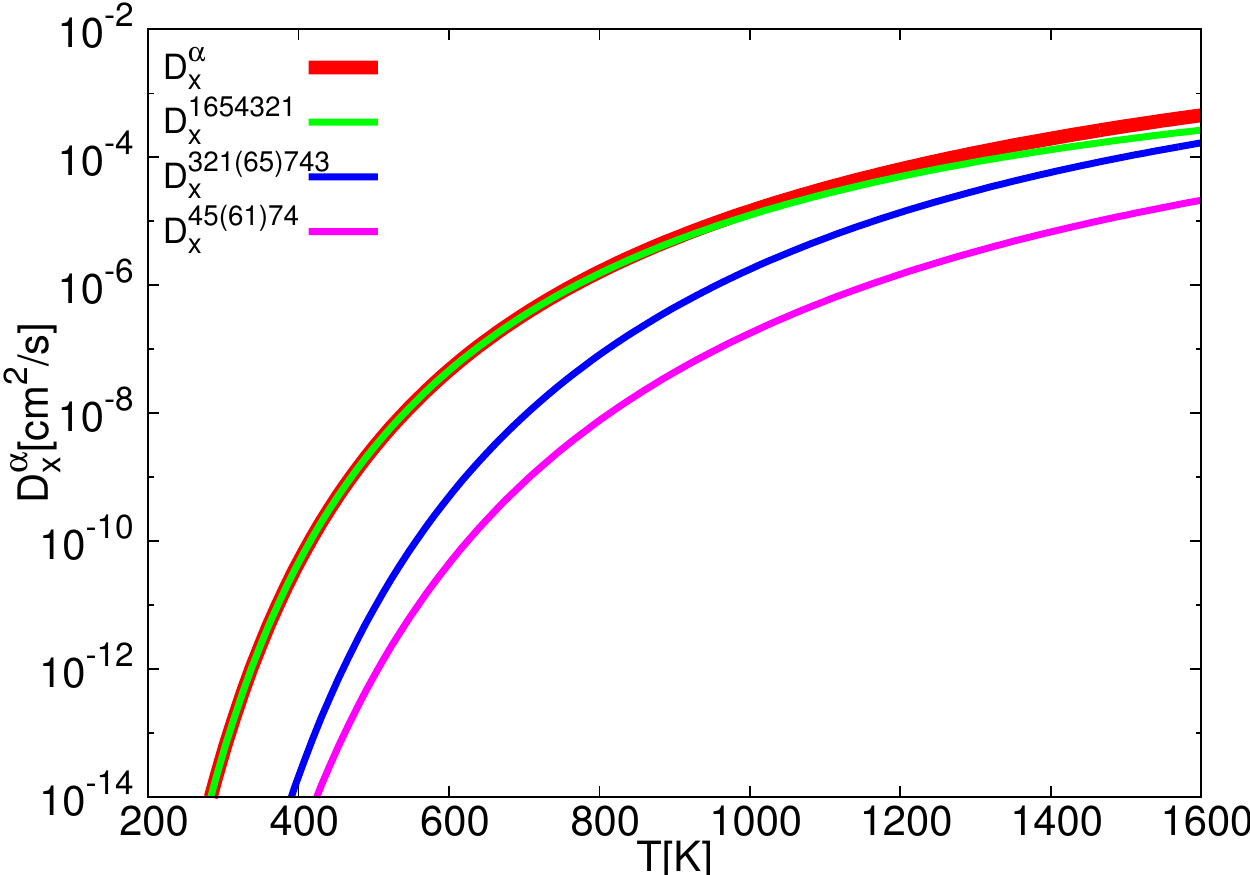}
\caption{Diffusion coefficient of Ga adatom in x direction over GaAs(001) $(4\times 4)\alpha$ reconstructed surface.\label{diffusion2x}}
\end{figure}
\begin{figure}
\includegraphics[width=7cm]{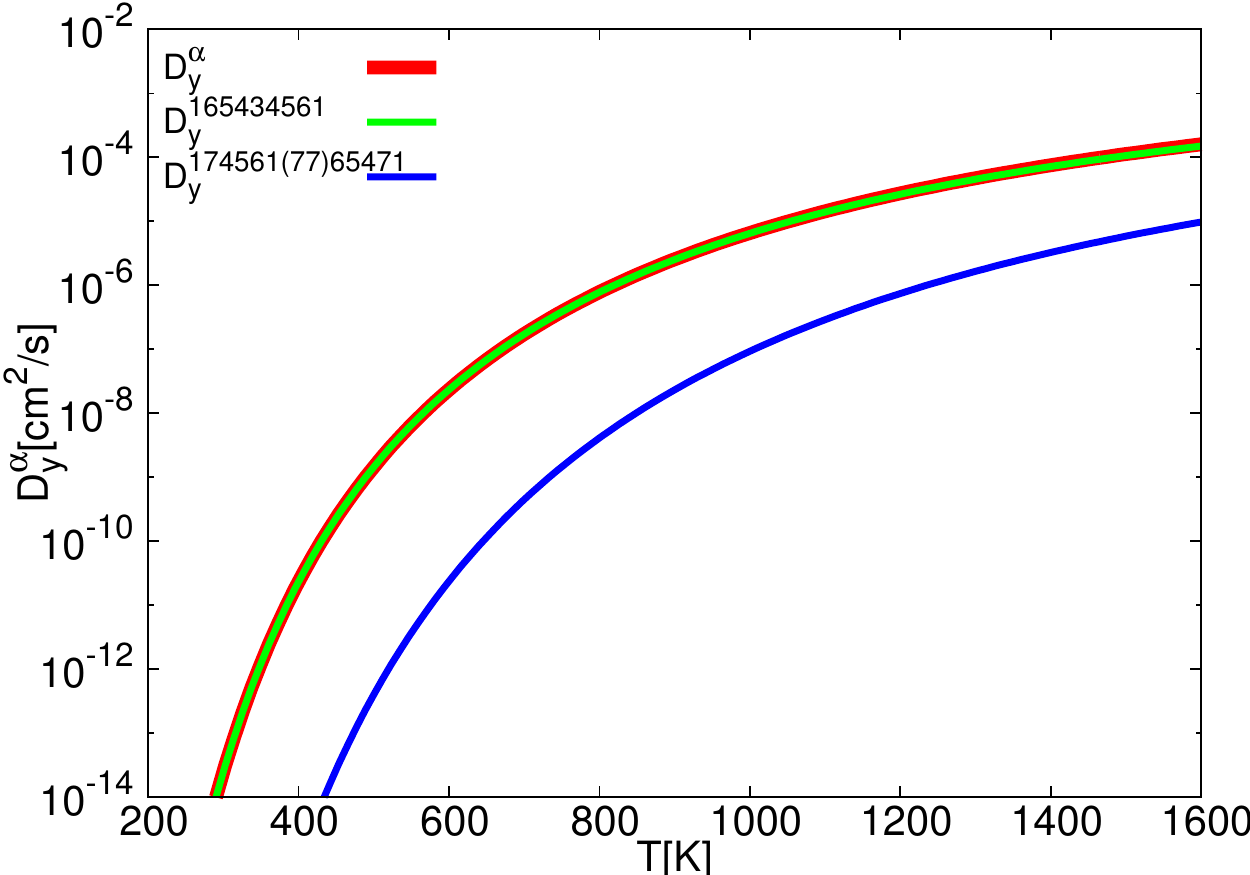}
\caption{Diffusion coefficient of Ga adatom in y direction over GaAs(001) $(4\times 4)\alpha$ reconstructed surface.\label{diffusion2y}}
\end{figure}
\begin{figure}
\includegraphics[width=7cm]{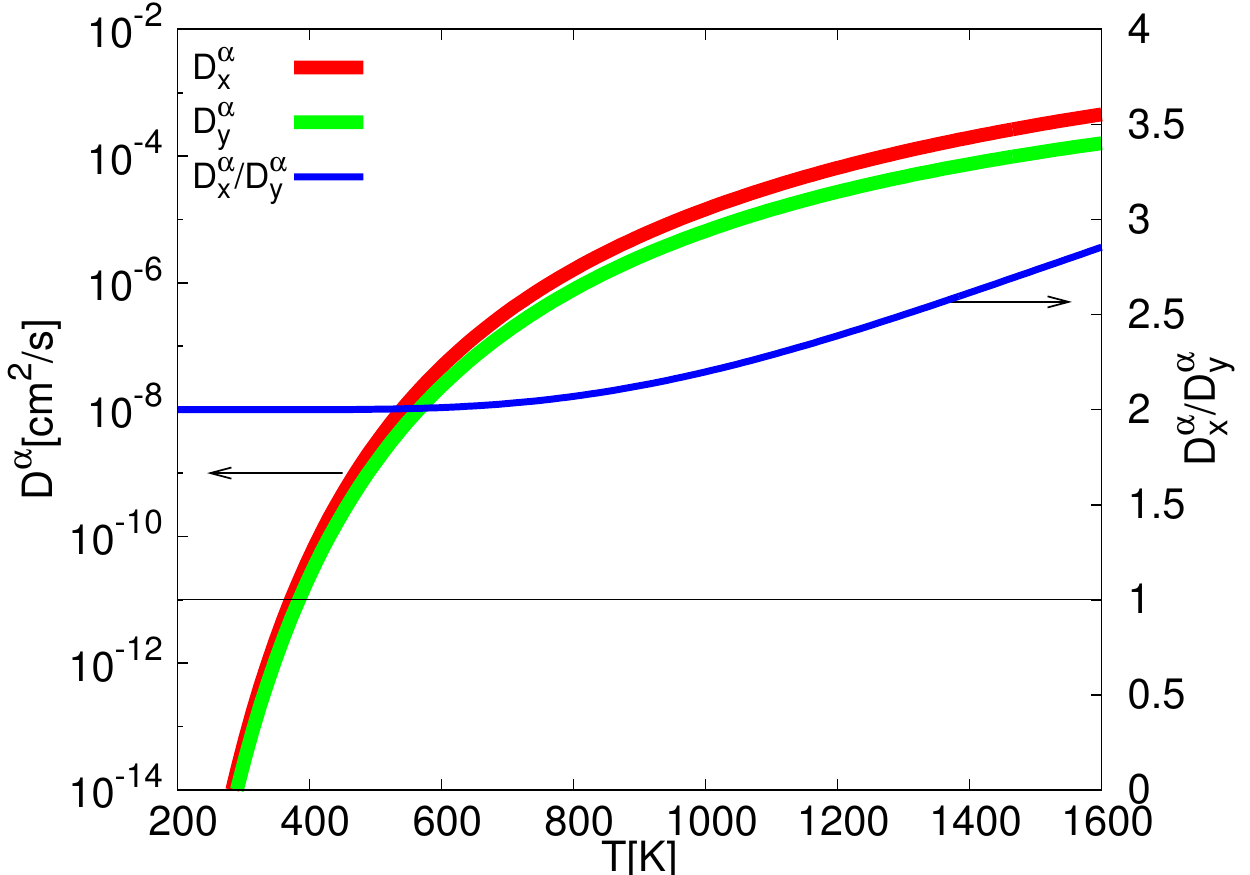}
\caption{Comparison of diffusion  coefficients in x and y directions for GaAs(001) $(4\times 4) \alpha$ reconstructed surface as  a temperature function in logarithmic scale. The scale of the anisotropy coefficient  is shown on the right.\label{compare2}}
\end{figure}
\begin{figure}
\begin{center}
\includegraphics[angle=0,width=7cm]{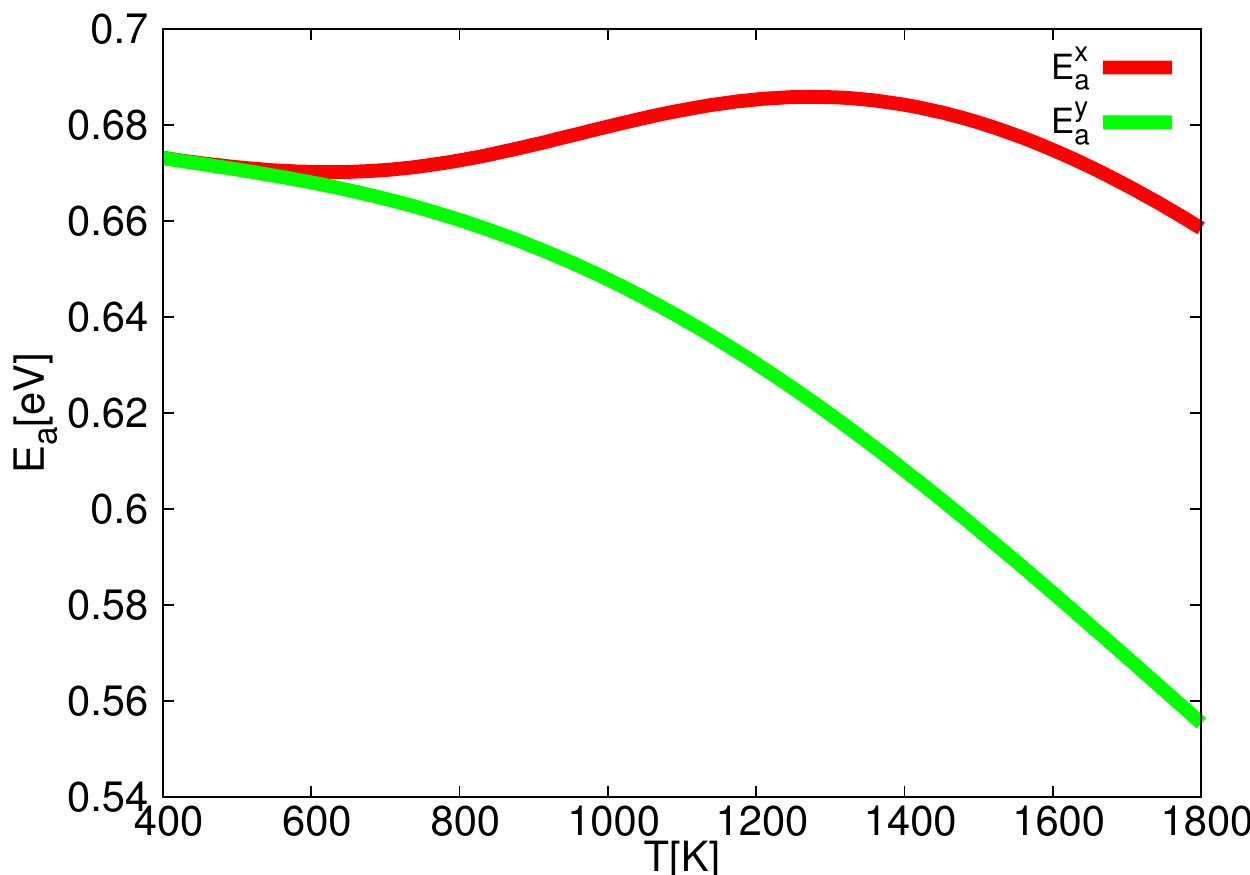}
\end{center}
\caption{Temperature dependence of the activation energies calculated with use of (\ref{ea}) for total diffusion coefficients $D^\alpha$ in both directions. \label{activation2}}
\end{figure}
\section{Diffusion coefficient for  $\rm Ga$ adatom  at  $\rm GaAs(001) - c(4 \times 4) \alpha$ reconstruction}
Let us take into regard $c(4\times 4)\alpha$  reconstruction  of the same GaAs(001)surface \cite{ohtake,roehl2}. The unit cell at this surface  is similar one to this in the $\beta$ reconstruction, but with slightly different   position of atoms at top layer. As a  result adsorption sites have  completely different locations and  energies \cite{roehl2}. Thus the symmetry of the  energy landscape for Ga adatom  is changed as can be seen in Fig \ref{lattice2}. Note that we have now
eleven different adsorption  sites  out of which nine (1,2,3,4a,4b,5a,5b,6a,6b) lie within trenches and two - namely 7a and 7b at dimer hills.  The lattice has inverse  symmetry    only with respect to the axis x.There is no symmetry  in the perpendicular  direction. The cross-section of  potential energy  landscape along two paths marked in Fig. \ref{lattice2} is presented in Fig. \ref{landscape2}.

Our calculations of diffusion coefficient  start with formulas for equilibrium probabilities
\begin{equation}
P_{eq}^{\alpha}=\frac{\rho e^{-\beta E_\alpha}}{\sum_{i=1}^7 e^{-\beta E_i}}
\label{peq2}
\end {equation}
Variational vector (\ref{vec}) is constructed with two independent phases for each lattice site. According to the lattice  symmetry $\phi_y^{3}=\phi_y^1=0$ and $\phi_x^{a}=\phi_x^{b}$, $\phi_y^{a}=-\phi_y^{b}$ for  sites $4,5,6$ and $7$. 
Finally  diffusion coefficients come out as 
\begin{equation}
D_{x}^\alpha=\frac{W_{14}W_{14}'\theta_1+W_{14}W_{41}' P_{eq}^4+2W_{14}'{\tilde W}_{41}\theta_4}{2W_{14}+W_{14}'+{\tilde W}_{14}} a^2
\end{equation}
\begin{equation}
D_{y}^\alpha=\frac{W_{14}W_{14}''P_{eq}^1 P_{eq}^1+W_{14}W_{43}P_{eq}^1 P_{eq}^4}{W_{14}P_{eq}^1+W_{14}''P_{eq}^1+W_{43}P_{eq}^4} a^2
\end{equation}
where
\begin{eqnarray}
\frac{1}{W_{14}P_{eq}^1}&=&\frac{1}{W_{16}P_{eq}^1}+\frac{1}{W_{65}P_{eq}^6}+\frac{1}{W_{54}P_{eq}^5}\\
\frac{1}{W_{14}'P_{eq}^1}&=&\frac{1}{2W_{17}P_{eq}^1+2W_{57}P_{eq}^5}+\frac{1}{2W_{74}P_{eq}^7}\nonumber\\
\frac{1}{\tilde{W}_{14}P_{eq}^1}&=&\frac{1}{2W_{43}P_{eq}^4}+\frac{1}{W_{32}P_{eq}^3}+\frac{1}{W_{21}P_{eq}^2}\nonumber \\
\frac{1}{W_{14}''P_{eq}^1}&=&\frac{1}{W_{17}P_{eq}^1+W_{57}P_{eq}^5+2W_{77}P_{eq}^7}+\frac{1}{W_{74}P_{eq}^7}\nonumber
\end{eqnarray}
In the above expression all particle transitions plotted in Fig. \ref{lattice2} were taken into account, but two of them, namely (57) and (17) were approximated by a simple sum as if they  both come out from the site (1). This is an approximation of very fast jumps between (1) and (5), i.e. much faster than both jumps (57) and (17) what is true in our case. Formulas obtained with this approximation are not far from the exact solution, but they have simpler structure and give an opportunity to analyze different diffusion  paths over the surface.  These paths are plotted in Figs \ref{path2x} and \ref{path2y}.

 It can be seen that there are three possible ways over which particle diffuses in x direction and two in direction y. Their contributions to the  global diffusion coefficient are presented in Figs \ref{diffusion2x} and \ref{diffusion2y}. The fastest paths  in both directions come along  trench and do not climb the hill. Diffusion along axis x happens mainly through two channels - first, more important one inside trench, and the  second one visits also dimer hills. Both these channels were identified in Ref. \onlinecite{roehl2} on the basis of energy differences for  consecutive jumps. The fastest  path in y direction comes through sites (16543)  and covers almost totally the value of  diffusion coefficient $D_y$. This last  diffusion channel  was not proposed  before as a possible diffusion path\cite{roehl2} whereas  it is important and interesting  example of particle way that avoids the lowest energetically site (now it is site 2).

An  anisotropy of the particle diffusion at the  reconstructed surface $(4\times4) \alpha$  is shown in Fig. \ref{compare}. Within  whole plotted range of temperatures diffusion is larger in direction x (parallel  to dimers) than in direction y (perpendicular to dimers). Blue dashed line shows ratio between $D_x$ and $D_y$ coefficients.
It starts from 2 at low temperature and increases to 3 for high temperatures. Thus comparing  these values with those calculated for the  surface  $(4\times 4) \beta$   in Fig. \ref{compare}  we conclude that the diffusion  anisotropy is higher at $\alpha$ surface reconstruction. It means that the  change  of surface symmetry affects symmetry of diffusion coefficient of single Ga particles. When we compare  activation barriers for diffusion along direction $x$ and  $y$ in Fig. \ref{activation2} we find the same situation as at the previously described surface, namely $E_A^x$ is larger than $E_A^y$.  We again can explain this fact through the number of possible diffusion ways.  Activation energies compared  for diffusion over surface $\alpha$ and $\beta$ are  lower for more complex structure of $\alpha$ symmetry in both directions.  We can understand such property as the dependence of the diffusion  coefficient on  larger number of  transition  rates in the $\alpha$ phase. As a result overall diffusion coefficient has relatively low activation energy value and low prefactor $\nu$.

\section{Conclusions}
 We have calculated diffusion coefficients for two versions of $c(4\times 4)$ reconstructed surface of GaAs(001)  $\beta$ and  $\alpha$ reconstructions. On using variational formula for the diffusion coefficient we were able to derive global expression for the elements of the diffusion coefficient tensor on the base of  the derived in ab-initio calculations energy landscape  of adsorption sites and  energy barriers \cite{roehl,roehl2}. In the used here coordinate system our diffusion tensor has only diagonal elements.
Diffusion along  direction parallel to dimer orientation ( x) is higher in all cases. However at low temperatures at $\beta$ surface phase diffusion is almost isotropic. The   anisotropy increases  with temperature and is higher for the  $\alpha$  surface phase. 

The structure of diffusion coefficient as a sum of different components allows for identification of different diffusion paths. Some of these paths agree with ones guessed from the observation of energy barrier for successive  jumps.  The method allows for finding also other paths not so obvious while  energy structure is studied. 
It appears that $\alpha$ structure is far more asymmetric that $\beta$ structure. 
 And also when we compare values of coefficients it comes out that diffusion over the surface with lower symmetry, $ (4 \times 4)\alpha $  is  of one order of magnitude lower that that for the surface of $(4 \times 4) \beta$ symmetry. It appears that  changes of surface structure  towards  system of lower symmetry  can be noted in the decrease of the diffusion process. Faster, more effective diffusion process at given temperature is characteristic for the surface $\beta$, the reconstruction of higher symmetry .

\section{Acknowledgement}
Research supported  by the National Science Centre(NCN) of Poland
(Grant NCN No. 2011/01/B/ST3/00526)

\end{document}